\documentclass[showpacs,twocolumn,preprintnumbers,amsmath]{revtex4}  

\usepackage{graphicx}
\usepackage{dcolumn}
\usepackage{bm}
\usepackage[dvipdf]{hyperref}
\usepackage{setspace}

\newcommand{\sro}{{$^{88}$}Sr}

\newcommand{\tripl}{$^1$S$_0$-$^3$P$_1$}
\newcommand{\singl}{$^1$S$_0$-$^1$P$_1$}

\draft

\begin{document}

\title{Long-lived Bloch oscillations with bosonic Sr atoms and application to gravity measurement at micrometer scale}

\author{G. Ferrari}
\author{N. Poli}
\author{F. Sorrentino}
\author{G. M. Tino}
\email{Guglielmo.Tino@fi.infn.it}

\affiliation{Dipartimento di Fisica and LENS, Istituto Nazionale Fisica Nucleare, Istituto
Nazionale Fisica della Materia, Polo Scientifico-Università di Firenze, 50019 Sesto Fiorentino,
Italy}

\begin{abstract}
We report on the observation of Bloch oscillations on the unprecedented time scale of several
seconds. The experiment is carried out with ultra-cold bosonic strontium-88 loaded into a vertical
optical standing wave. The negligible atom-atom elastic cross section and the absence of spin
makes $^{88}$Sr an almost ideal Bose gas insensitive to typical mechanisms of decoherence due to
thermalization and to external stray fields. The small size enables precision measurements of
forces at micrometer scale. This is a challenge in physics for studies of surfaces, Casimir
effects, and searches for deviations from Newtonian gravity predicted by theories beyond the
standard model.
\end{abstract}

 \maketitle

Quantum devices using ultracold atoms show extraordinary features. Atom interferometry is used for
precision inertial sensors \cite{peters99,gustavson00}, to measure fundamental constants
\cite{wicht02,clade06,stuhler03}, and testing relativity \cite{fray04}. The small size enables
precision measurements of forces at micrometer scale. This is a challenge in physics for studies
of surfaces, Casimir effects \cite{antezza05}, and searches for deviations from Newtonian gravity
predicted by theories beyond the standard model \cite{long05,smullin05}. Here we show that using
laser-cooled strontium atoms in optical lattices, persistent Bloch oscillations are observed for t
$\simeq$ 10 s, and gravity is determined with ppm sensitivity on micrometer scale. The
insensitivity to stray fields and collisions makes Sr in optical lattices, a candidate also for
future clocks \cite{takamoto05}, a unique sensor for small-scale forces with better performances
and reduced complexity compared to proposed schemes using degenerate Bose \cite{anderson98} or
Fermi \cite{roati04} gases. This improves the feasibility of new experiments on gravity in
unexplored regions.

The confinement of ultracold atoms in optical lattices, regular structures created by interfering
laser beams where the atoms are trapped by the dipole force, provides clean model systems to study
condensed-matter physics problems \cite{bloch05}. In particular, Bloch oscillations were predicted
for electrons in a periodic crystal potential in presence of a static electric field
\cite{bloch29} but could not be observed in natural crystals because of the scattering of
electrons by the lattice defects. They were directly observed using atoms in an optical lattice
\cite{raizen97}.

In our experiment, laser-cooled $^{88}$Sr atoms are trapped in a 1-dimensional vertical optical
lattice. The combination of the periodic optical potential and the linear gravitational potential
gives rise to Bloch oscillations at frequency $\nu_B$ given by

\[
\nu_B= \frac{m g \lambda_L}{2h}
\]

where $m$ is the atomic mass, $g$ is the acceleration of gravity, $\lambda_L$ is the wavelength of
the light producing the lattice, and $h$ is Planck's constant. Since both $\lambda_L$ and $m$ are
well known, the overall force along the lattice axis can be determined by measuring the Bloch
frequency $\nu_B$. In order to do a force measurement with given interrogation time, the atomic
wavefunction has to undergo a coherent evolution on the same time timescale. The most common
effects limiting the coherence time for ultracold atoms are perturbations due to electromagnetic
fields and atom-atom interactions. $^{88}$Sr is in this respect a good choice because in the
ground state it has zero orbital, spin and nuclear angular momentum that makes it insensitive to
stray electric and magnetic fields that otherwise need to be shielded. In addition, $^{88}$Sr has
remarkably small atom-atom interactions \cite{ferrari06}; this prevented so far the achievement of
Bose-Einstein condensation for this atom \cite{ferrari06,ido00} but becomes an important feature
in experiments where collisions lead to a loss of coherence limiting the measurement time and the
potential sensitivity.

\begin{figure}[t] \vspace{-0mm} \begin{center}
\hspace{-0mm}\includegraphics[width=0.45\textwidth,angle=0]{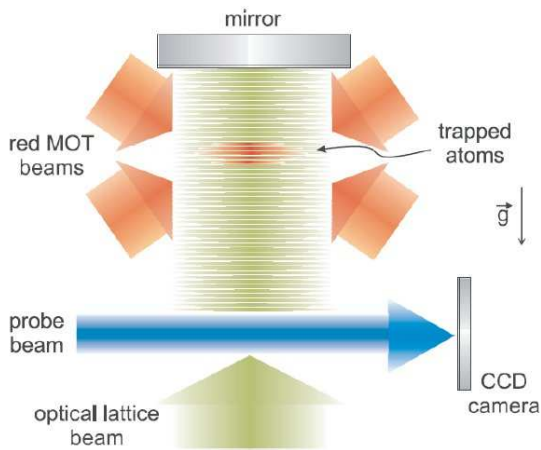}

\vspace{-0mm} \caption{\label{figura1} Simplified scheme of the apparatus used to observe Bloch
oscillations and to measure g: Sr atoms are laser cooled and trapped at a temperature of about 400
nK in a red magneto-optical-trap (MOT). The MOT laser beams are then switched-off and the atoms
are transferred in a vertical 1-dimensional optical lattice generated by a laser beam
retroreflected by a mirror; atoms are confined in series of layers at the maxima of the standing
wave by the dipole force. We measure the momentum distribution of the atoms, after the coherent
evolution in the potential given by the periodic potential plus gravity, by a time-of-flight
measurement, after a free fall of 12 ms, using a resonant probe laser beam and absorption imaging
on a CCD camera.}
\end{center}
\end{figure}

The experimental setup used in this work is schematically shown in Fig. \ref{figura1}. The method
used to produce ultracold Sr atoms was already described in \cite{poli05}. The experiment starts
with trapping and cooling $\sim 5 \times 10^7$ \sro\, atoms at 3 mK in a magneto-optical trap
(MOT) operating on the \singl blue resonance line at 461 nm. The temperature is then further
reduced by a second cooling stage in a red MOT operating on the \tripl narrow transition at 689 nm
and finally we obtain $\sim 5 \times 10^5$ atoms at 400 nK. After this preparation phase, that
takes about 500 ms, the red MOT is switched off and a one-dimensional optical lattice is switched
on adiabatically in 50 $\mu$s. The lattice potential is originated by a single-mode
frequency-doubled Nd:YVO4 laser ($\lambda_L$ = 532 nm) delivering up to 350 mW on the atoms with a
beam waist of 200 $\mu$m. The beam is vertically aligned and retro-reflected by a mirror producing
a standing wave with a period $\lambda_L$/2 = 266 nm. The corresponding photon recoil energy is
$E_R = h^2 /2m \lambda_L^2 = k_B \times 381$ nK. As expected from band theory \cite{ashcroft}, the
amplitude of the oscillation in momentum space decreases as the lattice depth is increased. This
suggests that in order to measure the Bloch frequency with maximum contrast the intensity of the
lattice laser should be reduced. On the other hand, reducing the intensity results in a loss in
the number of trapped atoms because of the smaller radial confinement. For this reason, we used a
lattice depth of 10 $E_R$. For a lattice potential depth corresponding to 10 $E_R$, the trap
frequencies are 50 kHz and 30 Hz in the longitudinal and and radial direction, respectively.
Before being transferred in the optical lattice, the atom cloud in the red MOT has a disk shape
with a vertical size of 12 $\mu$m rms. In the transfer, the vertical extension is preserved and we
populate about 100 lattice sites with $2 \times 10^5$ atoms with an average spatial density of
$\sim 10^{11}$ cm$^{-3}$. After letting the atoms evolve in the optical lattice, the lattice is
switched off adiabatically and we measure the momentum distribution of the sample by a
time-of-flight measurement, after a free fall of 12 ms, using a resonant probe laser beam and
absorption imaging on a CCD camera.

\begin{figure*}[t] \vspace{0mm} \begin{center}
\hspace{10mm}\includegraphics[width=1.0\textwidth,angle=0]{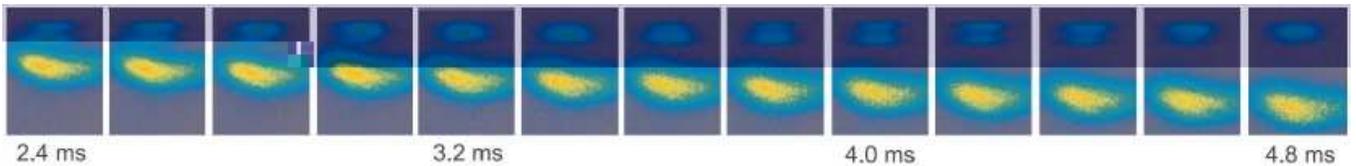}

\vspace{-0mm}  \caption{\label{figura2} Time-of-flight images of the atoms recorded for different
times of evolution in the optical lattice potential after switching-off the MOT. In the upper part
of each frame, the atoms confined in the optical lattice perform Bloch oscillations for the
combined effect of the periodic and gravitational potential. The average force arising from the
photon recoils transferred to the atoms compensates gravity. In the lower part, untrapped atoms
fall down freely under the effect of gravity.}
\end{center}
\end{figure*}

Fig. \ref{figura2} shows time-of-flight images of the atoms recorded for different times of
evolution in the optical lattice potential after switching-off the MOT. In the upper part of the
frames, the atoms confined in the optical lattice can be seen performing Bloch oscillations due to
the combined effect of the periodic and gravitational potential. The average force arising from
the photon recoils transferred to the atoms compensates gravity.

\begin{figure}[t] \vspace{-0mm} \begin{center}
\hspace{-0mm}\includegraphics[width=0.45\textwidth,angle=0]{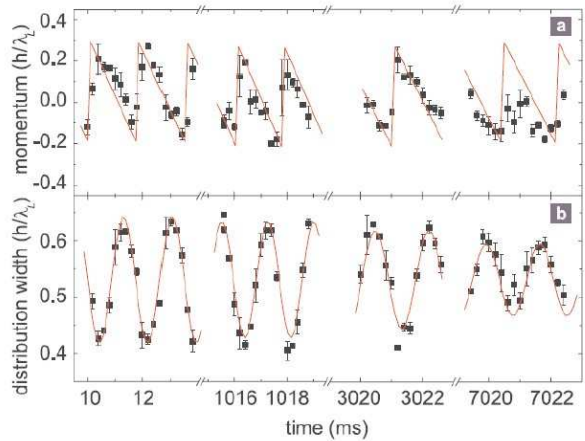}

\vspace{-0mm} \caption{\label{figura3} Bloch oscillation of \sro\, atoms in the vertical
1-dimensional optical lattice under the effect of gravity. Two quantities are extracted from the
analysis of the data: The average vertical momentum of the lower peak (a) and the width of the
atomic momentum distribution (b). From the fit of the data in b, a Bloch frequency $\nu_B$ =
574.568(3) Hz is obtained with a damping time $\tau \sim$ 12 s for the oscillations.}
\end{center}
\end{figure}

The images obtained by absorption imaging, as the ones shown in Fig. \ref{figura2}, are integrated
along the horizontal direction and fitted with the sum of two Gaussian functions. From each image,
two quantities are extracted that are reported in Fig. \ref{figura3} as a function of the
evolution time in the lattice: the first is the vertical momentum distribution of the lower peak,
that is shown in Fig. \ref{figura3}a. The second is the width of the atomic momentum distribution
(i.e. the second momentum of the distribution) which is shown in Fig. \ref{figura3}b. We find that
the latter is less sensitive against noise-induced perturbations to the vertical momentum. We
observed $\sim$ 4000 Bloch oscillations in a time $t$ = 7 s. During this time, about 8000 photon
momenta are coherently transferred to the atoms. Oscillations continue for several seconds and the
measured damping time of the amplitude is $\tau \sim $ 12 s. To our knowledge, the present results
for number of Bloch oscillations, duration, and the corresponding number of coherently transferred
photon momenta, are by far the highest ever achieved experimentally in any physical system. \\
From the measured Bloch frequency $\nu_B = 574.568(3)$ Hz we determine the gravity acceleration
along the optical lattice $g = 9.80012(5)$ ms$^{-2}$. The overall estimated sensitivity is $5
\times10^{-6} $ g and, neglecting the 500 ms preparation of the atomic sample, we have a
sensitivity of $4 \times10^{-5} $ g at 1 second. We expect that a sensitivity of $10^{-7} $ g can
be achieved using a larger number of atoms, and reducing the initial temperature of the sample.
Apart from collisional relaxation, which should contribute to decoherence on a minute timescale,
the main perturbation to quantum evolution is represented by vibrations of the retro-reflecting
mirror \cite{gilberto}. Minor contributions to decoherence may come from the axial momentum
dispersion of the lattice at $10^{-6}$ due to its radial extension.

\begin{figure}[t] \vspace{-0mm} \begin{center}
\hspace{-0mm}\includegraphics[width=0.45\textwidth,angle=0]{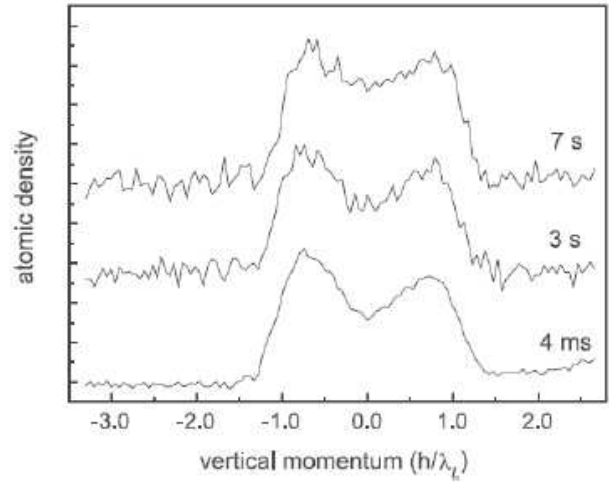}

\vspace{-0mm} \caption{\label{figura4} Vertical momentum distribution of the atoms at the Bragg
reflection recorded after an evolution time in the lattice of 4 ms, 3 s, and 7 s. Bloch
oscillations can be detected with a good contrast for times as long as 7 s with more than $10^4$
atoms. At longer times the signal-to-noise ratio is degraded due to loss of atoms induced by
collisions with the background vapor pressure. For a better visibility the vertical scale was
rescaled among the three graphs.}
\end{center}
\end{figure}

Fig. \ref{figura4} illustrates the loss of visibility of the two-component momentum distribution
at the Bragg reflection. The persistent visibility of the bimodal distribution on more than
$10^{4}$ atoms directly reflects the single atom coherent evolution on seconds timescale, while
the reduction in the signal-to-noise ratio is due to the 5 s $e^{-1}$ lifetime of the sample
limited essentially by residual background vapor pressure in the MOT chamber. It is worth noting
that the 5 s lifetime is not inconsistent with the observed 12 s coherence time: the atoms that
collide with the room temperature background vapor are ejected from the $\mu$K deep lattice
potential and do not contribute to the measured signal.

The micrometric spatial extension of the atomic cloud in the vertical direction, and the
possibility to load it into the optical potential at micrometric distance from a surface, makes
the scheme we demonstrated particularly suitable for the investigation of forces at small spatial
scales. The possibility of investigating the gravitational force at small distances by atomic
sensors was proposed in \cite{tino01bis}, discussed in detail in \cite{dimopulos03}, and
preliminarly demonstrated in \cite{haber05}. Deviations from the Newtonian law are usually
described assuming a Yukawa-type potential

\[
V(r)=-G\frac{m_1m_2}{r}(1+\alpha e^{-r/\lambda})
\]

where $G$ is Newton's gravitational constant, $m_1$ and $m_2$ are the masses, $r$ is the distance
between them. The parameter $\alpha$ gives the relative strength of departures from Newtonian
gravity and $\lambda$ is its spatial range. Experiments searching for possible deviations have set
bounds for the parameters $\alpha$ and $\lambda$. Recent results using microcantilever detectors
lead to extrapolated limits $\alpha \sim 10^4$ for $\lambda \sim 10$ $ \mu$m and for distances
$\sim 1$ $\mu$m it was not possible to perform direct experiments so far \cite{long05,smullin05}.

The small size and high sensitivity of the atomic probe allows a direct, model-independent
measurement at distances of a few $\mu$m from the source mass with no need for modeling and
extrapolation as in the case of macroscopic probes. This allows to directly access unexplored
regions in the $\alpha-\lambda$ plane. Also, in this case quantum objects are used to investigate
gravitational interaction.

Our results indicate that our Sr atoms when brought close to a thin layer can be used as probe for
the gravitational field generated by the massive layer\cite{kuhr01}. If we consider, in fact, a
material of density $\rho$ and thickness $d$, the added acceleration of gravity in proximity of
the source mass is $a = 2\pi G\rho d$ so that when $d \sim 10$ $\mu$m and $\rho \simeq$ 10
g/cm$^{3}$) as for tungsten crystals the resulting acceleration is $a \sim 4 \times 10^{-11}$
ms$^{-2}$. Measuring $\nu_B$  at a distance $\sim 4$ $\mu$m away from the surface
\cite{carusotto05} would allow to improve the constraint on $\alpha$ by two orders of magnitude at
the corresponding range $\lambda \sim 4$ $\mu$m. Spurious non-gravitational effects (Van der
Waals, Casimir forces), also present in other experiments, can be reduced by using an electrically
conductive screen and performing differential measurements with different source masses placed
behind it. Moreover, by repeating the same measure with the 4 stable isotopes (3 bosons, 1
fermion, with atomic mass ranging from 84 to 88), hence modulating the gravity potential through
the probe mass instead of the source mass, we can further discern among gravitational and QED
forces.

In conclusion, we observed persistent Bloch oscillations of weak-interacting bosonic Sr atoms in a
vertical optical lattice for a time longer than 7 s, with more than 8000 photon momenta coherently
transferred to the atoms. In addition to the intrinsic interest of the observed effect, these
results can be important for different experiments as, for example, in the measurement of
fundamental constants. The small size and high sensitivity of the new atomic sensor are important
for the investigation of small spatial-scale forces as in atom-surface interactions,
surface-induced decoherence, Casimir-Polder interaction, and for the search of recently predicted
deviations from the Newtonian gravitational law at the micrometer scale. Crucially enough, when
compared with other proposals using Bose-Einstein condensates and degenerate Fermi gases as probe,
our scheme is not affected by collisional and mean-field degrading effects. This enables us to
reach much longer observation times and higher sensitivities. In addition, our atoms are
insensitive to stray electric and magnetic fields and an all-optical procedure is used for the
cooling and the confinement. The preparation of the ultracold atom sample takes less than 0.5 s,
which is negligible with respect to measurement duration, and much faster the typical tens of
seconds needed for a cycle of evaporative and sympathetic cooling when using degenerated gases.

We thank M. Artoni and G. Oppo for a critical reading of the manuscript, G. Saccorotti for the
seismic noise measurements, and R. Ballerini, M. De Pas, M. Giuntini, A. Hajeb for technical
assistance. G. M. Tino acknowledges stimulating discussions with E. A. Cornell. This work was
supported by Istituto Nazionale di Fisica Nucleare, LENS, Ente Cassa di Risparmio di Firenze.

\end{document}